\documentclass[%
 reprint,
 amsmath,amssymb,
 aps,
]{revtex4-2}

\usepackage{graphicx}
\usepackage{dcolumn}
\usepackage{bm}

\newcommand{\be}{\begin{equation}}
\newcommand{\ee}{\end{equation}}
\newcommand{\ba}{\begin{eqnarray}}
\newcommand{\ea}{\end{eqnarray}}

\begin{document}

\preprint{APS/123-QED}

\title{Coarse-graining of the perplexity for the spatial distribution of molecules}

\author{Lili Xu$^{1}$}
\author{Chi Zhang$^{2}$}
\author{Soho Oyama$^{1}$}
\author{Manabu Machida$^{2,3}$}\email{machida@hiro.kindai.ac.jp}
\author{Tomoaki Kahyo$^{1,4}$}
\author{Mitsutoshi Setou$^{1,2,4}$}

\affiliation{
${^1}$ Department of Cellular and Molecular Anatomy, Hamamatsu University School of Medicine, Hamamatsu 431-3192, Japan\\
${^2}$ Department of Systems Molecular Anatomy, Institute for Medical Photonics Research, Preeminent Medical Photonics Education \& Research Center, Hamamatsu University School of Medicine, Hamamatsu 431-3192, Japan\\
${^3}$ Department of Informatics, Faculty of Engineering, Kindai University, Higashi-Hiroshima 739-2116, Japan\\
${^4}$ International Mass Imaging Center, Hamamatsu University School of Medicine, Hamamatsu 431-3192, Japan
}

\date{\today}

\begin{abstract}
Biological tissue consists of various molecules. Instead of focusing on a particular molecule, we consider the Shannon entropy which is calculated from the abundance of different molecules at each spot in the tissue. The spatial distribution of the Shannon entropy is of interest. In this paper, we first obtain the heat map of perplexity, whose logarithm is the entropy. To characterize the spatial variety of molecules, we propose a scalar $k$ that is concerned with the coarse-graining of the perplexity heat map. To verify the usefulness of the number, experiments with MSI (mass spectrometry imaging) were performed for mouse kidneys. We found that $k$ has large values in the renal pelvis area, cortex area, veins, and arteries in the mouse kidney, whereas fractal dimensions fail to distinguish those regions.
\end{abstract}

\maketitle


\section{Introduction}
\label{intro}

Among different molecules in biological tissue, quite often a target molecule is focused on and the position-dependence of the abundance of the target molecule is investigated. An alternative approach to study the tissue with various molecules is to obtain a characteristic quantity such as the Shannon entropy and study its position-dependence. In this paper, aiming at characterizing  biological tissues from the viewpoint of the spatial distribution of the entropy, we perform coarse-graining.

Entropy has been used to analyze information of a medium \cite{Frigg11}. For classical and quantum gases, fractal dimensions and their relation to entropy were investigated \cite{Buyukkilic-Demirhan93}. The Shannon entropy was used for avoided crossings in quantum chaos \cite{Park-etal18}. The concept of the Shannon entropy was used for kinetics of colloidal particles \cite{Wu-etal09}. See also \cite{Jaynes80, Grandy80}. Entropy was used for the study of quantum many-body systems \cite{Tarighi-etal22}. An extension of the Shannon entropy was explored for biological diversity \cite{Ricotta03}.

To study biological tissue from informational point of view, heat maps of the Shannon entropy calculated from the abundance of molecules in each spot has been investigated \cite{Aoyagi-etal22,Xu-etal23}. In this paper, characterization of the position dependence of the Shannon entropy is explored.

Mass spectrometry imaging (MSI) is a technique with ionizing molecules in a sample to provide both molecule species and positions of those molecules in the sample \cite{Shimma-etal07,Chughtai-Heeren10,Spengler15}. The resolution of an image is of the order of micrometer or subnanometer. At each spot of the sample, hundreds of $m/z$ peaks are obtained.

In mass spectrometry, the Shannon entropy was calculated for $m/z$ spectra 
\cite{Ferrige-Seddon-Jarvis91,Reinhold-Reinhold92,Broersen-etal08,Abdelmoula-etal14}. See \cite{Kaltashov-Eyles05,Aoyagi09} for more details. Recently, the binning effect of the Shannon entropy \cite{Madiona-etal19}, the relation between the Shannon entropy of mass spectra and molecules such as peptides and proteins \cite{Li-etal21}, and a data-targeted extraction method for metabolite annotation \cite{Zheng-etal22} were investigated.

The Shannon entropy has been viewed as a physical quantity which gives information how molecules spatially vary. In \cite{Aoyagi-etal22}, the Shannon entropy has been revisited to capture information from all peaks in mass spectra over the sample and a method based on the information entropy (Shannon entropy) for time-of-flight secondary ion mass spectrometry (TOF-SIMS) was proposed and it was shown that without peak identification the spatial distribution (heat maps) of the Shannon entropy of spectra indicates differences in materials and changes in the conditions of a material in a sample. The spatial distribution of the Shannon entropy was also studied for the matrix-assisted laser desorption/ionization (MALDI) MSI and a method to select candidate peaks was proposed \cite{Xu-etal23}.

In this study, we derive perplexity from entropy, and develop an approach to visualize the spatial and mass-spectral diversities by coarse-graining. To this end, a slope $k$ (see below) is introduced.

The remainder of the paper is organized as follows. In Sec.~\ref{perpl}, we introduce perplexity and define $k$. Sec.~\ref{exp}, experimental results for MSI are shown. Fractal dimensions are considered in Sec.~\ref{frac}. Section \ref{discuss} is devoted to discussion. The conclusions are given in \ref{concl}.

\section{Perplexity and coarse-graining}
\label{perpl}

We begin by introducing the Shannon entropy using intensities in the mass spectrum \cite{Aoyagi-etal22,Xu-etal23}. Let $n$ be the number of intensities in the mass spectrum. In MSI, different intensities appear as a function of $m/z$ at each pixel on the sample. Examples of such peaks are shown in Fig.~\ref{peaks} for the mouse kidney of Sample 4-2 (see below). We call nonzero intensities as peaks. Each peak corresponds to an ion. Here, the dimensionless unit $m/z$ is the ratio of the mass of an ion divided by the unified atomic mass unit ($1$ Dalton) and the absolute charge number (the absolute value of an integer number of elementary charges that an ion has gained or lost). Let $p_i$ be the $i$th intensity in the mass spectrum which is normalized as $\sum_{i=1}^np_i=1$. In Fig.~\ref{peaks}, the numbers of peaks are $558$, $478$, $655$ for Points A,B,C, respectively.

We draw $x$- and $y$-axes on the image. At pixel $(x,y)$, we define the Shannon entropy $H(x,y)$ as
\begin{equation}
H(x,y)=-\sum_{i=1}^np_i(x,y)\log_2{p_i(x,y)}.
\end{equation}
After plotting $H(x,y)$, a way to study the structure of $H(x,y)$ in the neighborhood of the point $(x,y)$ is to consider coarse-graining. As described below, a computable scalar can be obtained with the coarse-graining process for perplexity, which is introduced below, rather than the direct application of coarse-graining to the heat map $H(x,y)$.

\begin{figure}[htbp]
\centering
\includegraphics[width=0.45\textwidth]{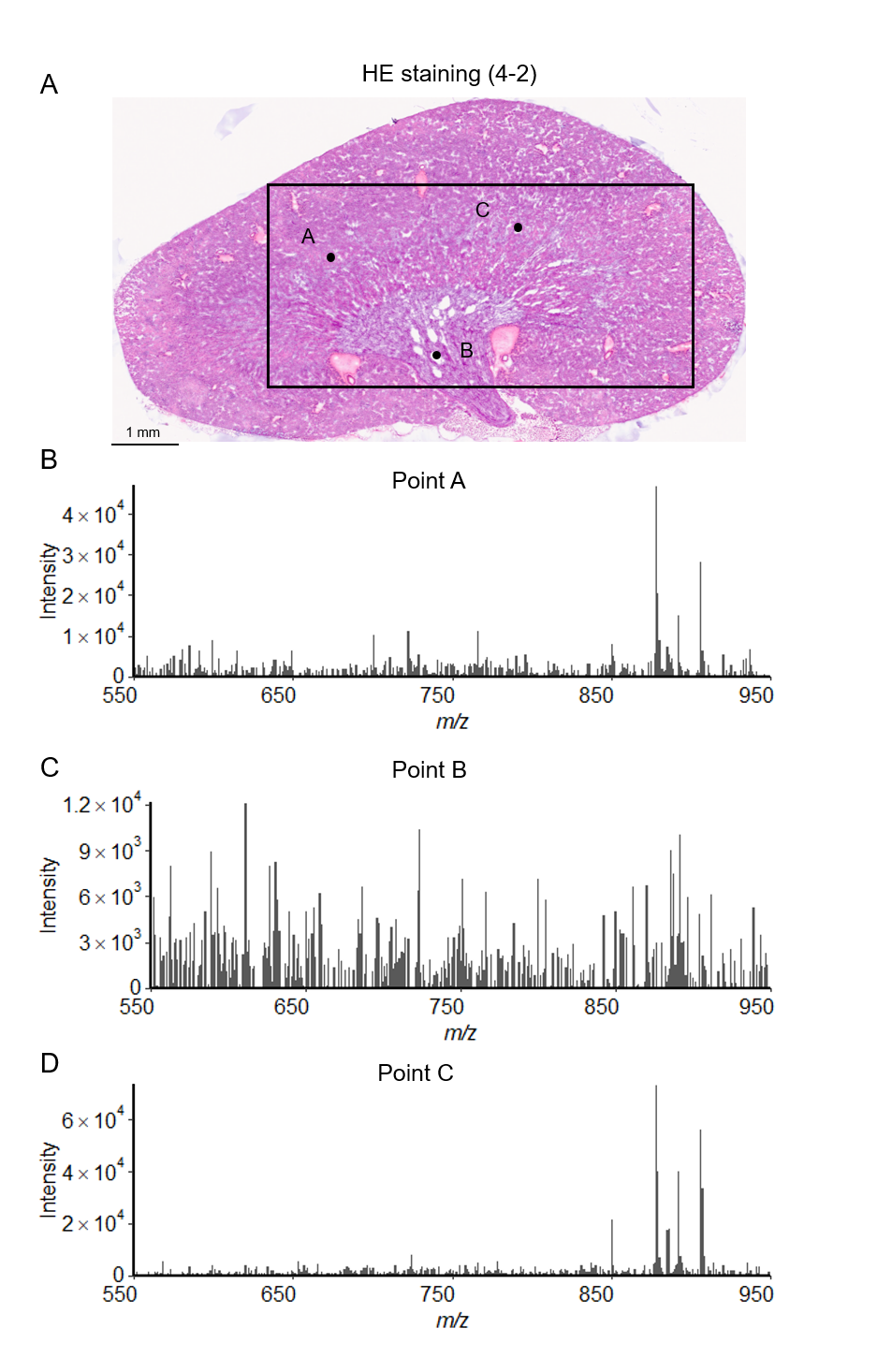}
\caption{\label{peaks}
A mouse kidney (Sample 4-2) (color online) and examples of peaks on mass spectra at three different spots in the kidney.
}
\end{figure}

Let us consider perplexity, which is an index to measure diversity. See \cite{Jost06} for an application of perplexity in biology. Perplexity ($PP$) is defined as \cite{Nelson14}
\begin{equation}
PP(x,y)=2^{H(x,y)}.
\end{equation}
Suppose that we randomly pick a peak in the mass spectrum with the probability given by the relative height of the peak. Then we consider the number of tries that is necessary for the particular peak to be selected. Indeed, the expected number of trials is the reciprocal of the probability for the peak. The perplexity is the weighted geometrical mean of the expected numbers of trials for all peaks in the mass spectrum. In this way, the perplexity provides a measure of diversity of peaks, which correspond to different molecules. Since $0\le H(x,y)\le \log_2{n}$, we have $1\le PP(x,y)\le n$.

Let us consider coarse-graining for MSI. We select a square region on the image and divide the square into $N(\varepsilon)$ sub-squares with side length $\varepsilon$. Treating these $N(\varepsilon)$ sub-squares as new pixels, relative peak intensities in the pixel at $(x,y)$ can be expressed as $p_i^{(\varepsilon)}(x,y)$. We take averages in mass spectra for coarse-grainned pixels. If the pixel size gets doubled by coarse-graining, we have
$p_i^{(2\varepsilon)}(x,y)=
\frac{1}{4}\left[p_i^{(\varepsilon)}(x_1,y_1)+p_i^{(\varepsilon)}(x_2,y_2)
+p_i^{(\varepsilon)}(x_3,y_3)+p_i^{(\varepsilon)}(x_4,y_4)\right]$,
where $(x_1,y_1)=(x,y)$, $(x_2,y_2)=(x+\varepsilon,y)$, $(x_3,y_3)=(x,y+\varepsilon)$, and $(x_4,y_4)=(x+\varepsilon,y+\varepsilon)$. We note that the pixels of size $\varepsilon$ at $(x_j,y_j)$ ($j=1,2,3,4$) are contained in the pixel of size $2\varepsilon$ at $(x,y)$. Thus, the perplexity depends on $\varepsilon$ and we can write $PP^{(\varepsilon)}(x,y)$.

In addition to $PP^{(\varepsilon)}(x,y)$ itself, we consider how $PP^{(\varepsilon)}(x,y)$ behaves as $\varepsilon$ varies by coarse-graining. We define $k(x,y)$ as
\begin{equation}
k(x,y)=\lim_{\varepsilon\to0}\frac{PP^{(\varepsilon)}(x,y)}{\ln\varepsilon}.
\label{kdef}
\end{equation}
This $k(x,y)$ contains the information on how $H(x,y)$ behaves in the neighborhood of the point $(x,y)$. Since we have found a linear dependence between $\ln\varepsilon$ and the perplexity (see below), the calculation of the limit in (\ref{kdef}) is feasible for experimental data. Indeed, (\ref{kdef}) means as a power-law behavior of $e^{PP}$ for small $\varepsilon>0$ as
\begin{equation}
e^{PP}\sim\varepsilon^k.
\label{PPeps}
\end{equation}
Below, we will see the relation (\ref{PPeps}) holds for experimental data. In terms of the Shannon entropy, the relation (\ref{PPeps}) implies a power-law dependence of the double-exponential of $H$ on $\varepsilon$:
\begin{equation}
\exp\left(e^H\right)\sim\varepsilon^k.
\end{equation}
Thus, $k$ characterizes the local behavior of the spatial distribution of $H$.

\section{Experiments}
\label{exp}

\subsection{Setup}
\label{setup}

We used C57BL/6 J female mice, two 4-month-old mice kidneys (Samples 4-1 and 4-2) for MALDI MSI observation (a high-resolution microscopic imaging mass spectrometer, iMScope). Sample preparation steps including mouse sacrifice, sample pre-reservation, and matrix spraying followed our previous paper \cite{Xu-etal23}. The experimental conditions were: in negative ion mode, $m/z$ range between $550$ and $950$, laser strength of $45\%$, and number of irradiations of $100$. All experiments in this study were performed in compliance with the licensing instructions from the Institutional Animal Care and Use Committees of Hamamatsu University, School of Medicine, Japan (permission code: 2015028).

For Sample 4-1, $n=3246$. For Sample 4-2, $n=3584$.

\subsection{Perplexity heat maps}
\label{heatmap}

\begin{figure}[htbp]
\centering
\includegraphics[width=0.4\textwidth]{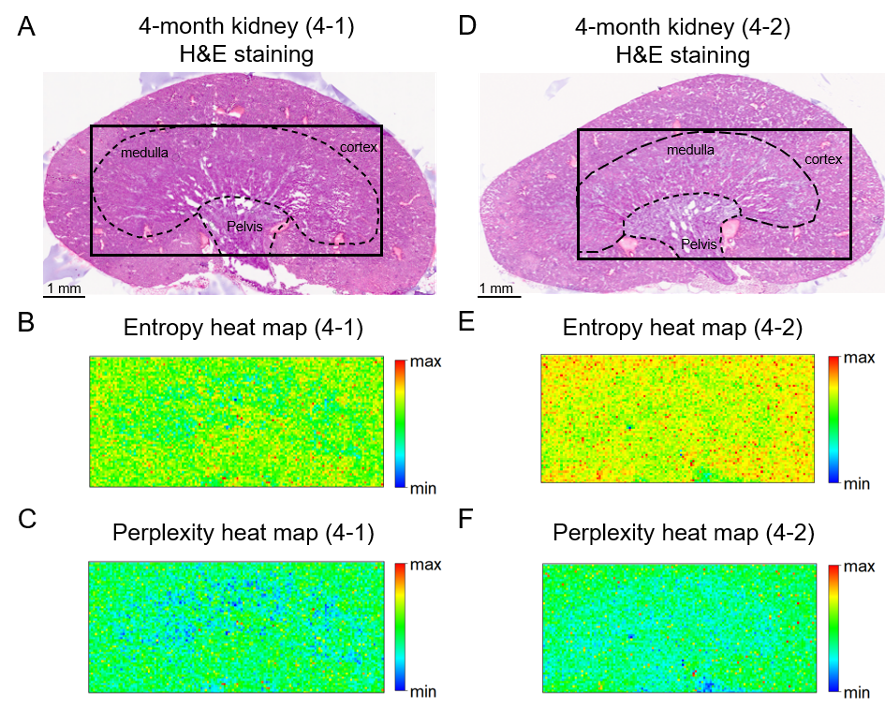}
\caption{\label{fig3}
(Color online) Entropy and perplexity heat maps on kidney MSI data. (A, D) H\&E staining of kidneys (Samples 4-1 and 4-2) of $4$-month mice and ROIs (marked by rectangles). (B, E) Entropy heat maps in ROIs. In the ROI of Sample 4-1, the entropy changed between $5.0$ and $9.6$; In the ROI of Sample 4-2, the entropy changed between $7.2$ and $9.7$. (C, F) Perplexity heat maps in ROIs. In the ROI of Sample 4-1, the perplexity changed between $149.0$ and $846.5$; In the ROI of Sample 4-2, the perplexity changed between $32.2$ and $755.3$. Scale bar: $1\,{\rm mm}$.
}
\end{figure}

Firstly, we select a region of interest (ROI). Figure \ref{fig3}A,D shows the anatomical structure through H\&E (hematoxylin-eosin) staining. We calculated the Shannon entropy and perplexity on each spot of the kidney MSI data and plotted heat maps. In Fig.~\ref{fig3}B,C (Sample 4-1) and Fig.~\ref{fig3}E,F (Sample 4-2), there are differences between entropy and perplexity heat maps. Moreover sample dependencies can be seen in the comparisons between Fig.~\ref{fig3}B and Fig.~\ref{fig3}E and between Fig.~\ref{fig3}C and Fig.~\ref{fig3}F. In Fig.~\ref{fig3}C, blue and green spots (i.e., low and middle perplexity spots) dominate in the medulla and cortex areas, respectively. In Fig.~\ref{fig3}E, the pelvis and medulla areas (green) (i.e., middle perplexity areas) can be distinguished from the cortex (yellow) (i.e., high perplexity spots). For both samples of 4-1 and 4-2, the medulla area has lower entropy values than the cortex area.

\subsection{The slope $k$}
\label{expk}

\begin{figure}[htbp]
\centering
\includegraphics[width=0.45\textwidth]{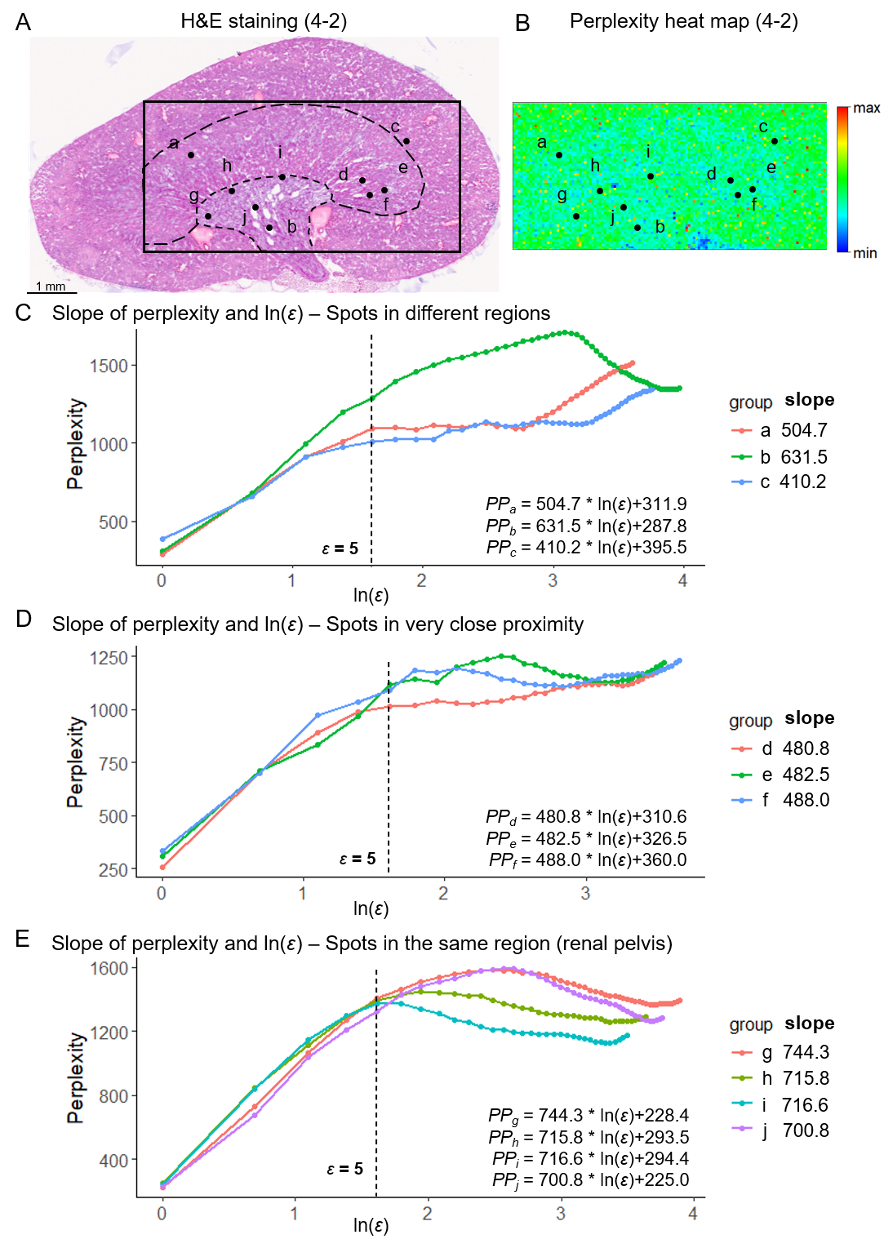}
\caption{\label{fig4}
(Color online) (A, B) The H\&E staining and perplexity heat map of the 4-month mouse kidney (Sample 4-2) and spots selected for demonstrating slope calculations. The range of the perplexity was between $32.2$ and $755.3$. (C) Semi-log plots of $\ln{\varepsilon}$ and the perplexity. For spots a, b, and c, slopes were $504.7$, $631.5$, and $410.2$, respectively. (D) For spots d, e, and f, slopes were $480.8$, $482.5$, and $488.0$, respectively. (E) For spots g, h, i, and j, slopes were $744.3$, $715.8$, $716.6$ and $700.8$, respectively. Scale bar: $1\,{\rm mm}$.
}
\end{figure}

To investigate the spatial distribution of perplexity, we picked several points on the image for the kidney of Sample 4-2 (Fig.~\ref{fig4}A, B). We found a linear trend in the semi-log plot of $\ln{\varepsilon}$ and perplexity near the origin. In Fig.~2, the unit of $\varepsilon$ is the length of the original pixels. Different spots (a, b, and c) were chosen and the linear dependence was found at each spot with different slopes (a: 504.7, b: 631.5, and c: 410.2) (Fig.~\ref{fig4}C). Spots in the proximity (d, e, and f) showed similar slopes (d: 480.8, e: 482.5, and f: 488.0) (Fig.~\ref{fig4}D). Spots in the same region (g, h, i, and j) presented similar slopes (g: 744.3, h: 715.8, i: 716.6 and j: 700.8) (Fig.~\ref{fig4}E). Thus, $k$ can be experimentally determined.

\begin{figure}[htbp]
\centering
\includegraphics[width=0.45\textwidth]{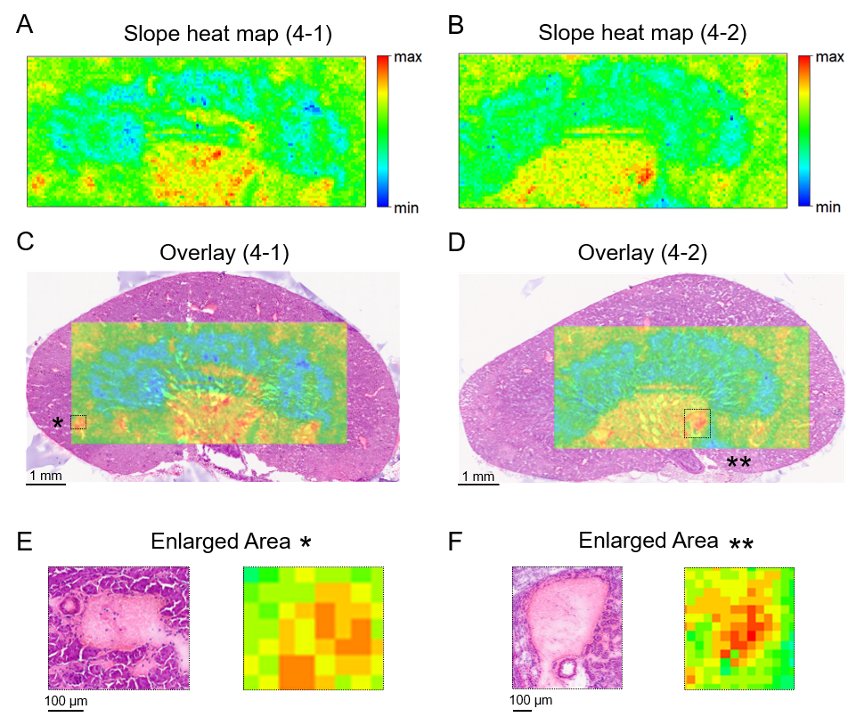}
\caption{\label{fig5}
(Color online) Heat maps of the slope $k$ are shown with overlays with H\&E staining. (A, B) Heat maps of $k$ for Samples 4-1 and 4-2. In A,B, $[{\rm min},{\rm max}]$ is $[171.1,1105.4]$ and [$104.4,902.6$], respectively. (C, D) Overlays of the heat map and H\&E staining of the kidneys. Scale bar: $1\,{\rm mm}$. (E, F) Enlarged areas for the H\&E staining and heat map of $k$. Scale bar: $10\,\mu{\rm m}$.
}
\end{figure}

Figure \ref{fig5}A,B shows heat maps of $k(x,y)$. We found that the renal pelvis area has large $k$ (red and yellow spots), while $k$ is small (blue and green spots) in the medulla area (Fig.~\ref{fig5}C,D). There were green and yellow spots in the cortex area. We also found pixels of large $k$ in the areas denoted by * and ** in Fig.~\ref{fig5}C,D.  As examples, we compare H\&E images and heat maps in these rectangular areas. The magnified figures are shown in Fig.~\ref{fig5}E,F. We discovered that large $k$ appears at the vein and artery.

\section{Fractal dimensions}
\label{frac}

Here we take the unit of $\varepsilon$ to be the side length of original pixels.  Let $N_0$ be the number of original pixels in the ROI. The number of pixels after binning is $N(\varepsilon)=N_0/\varepsilon^2$. Fractal dimension $D_0$ is calculated as $D_0=-\lim_{\varepsilon\to0}\ln{N(\varepsilon)}/\ln{\varepsilon}$ \cite{Kak20}. From the definition, we see that in this case exactly $D_0=2$. For given $\varepsilon$, let us introduce
\begin{equation}
P_i(\varepsilon)=\frac{\sum_{(x,y)\in\,i{\rm th}\,{\rm pixel}}H(x,y)}{\sum_{(x,y)\in\,{\rm ROI}}H(x,y)},
\end{equation}
where $H(x,y)$ is the Shannon entropy which is computed using original pixels. The number of pixels depends on $\varepsilon$; the number decreases as $\varepsilon$ becomes large.

Similar to $D_0$, the information dimension $D_1$ and correlation dimension $D_2$ \cite{Renyi59,Falconer03} are given by 
$D_1=\lim_{\varepsilon\to0}\sum_{i=1}^{N(\varepsilon)}P_i(\varepsilon)\ln{P_i(\varepsilon)}/\ln{\varepsilon}$,
$D_2=\lim_{\varepsilon\to0}\ln\left[\sum_{i=1}^{N(\varepsilon)}P_i(\varepsilon)^2\right]/\ln{\varepsilon}$. In general, we have $D_q=\frac{1}{q-1}\lim_{\varepsilon\to0}\left(\ln{Z_q(\varepsilon)}\right)/(\ln{\varepsilon})$, where $Z_q(\varepsilon)=\sum_{i=1}^{N(\varepsilon)}P_i(\varepsilon)^q$.

In addition to the fractal dimension $D_0$, the information dimension $D_1$ and the correlation dimension $D_2$ were calculated at different points in the images. We found $D_1=2.0$ and $D_2=2.0$ for both kidneys. This means that fractal nature is not found for $q=0,1,2$.

\section{Discussion}
\label{discuss}
We found that the relation $e^{PP}\sim\varepsilon^k$ in (\ref{PPeps}) holds for experimental data. Although this relation stems from statistical nature of the molecules, it is still an open problem how $k$ reflects the spatial distribution of molecules in the sample. The physical and biological reasons of the power-law behavior need to be clarified in the future.

We note that $k(x,y)$ is positive even when the Shannon entropy for pixels in the neighborhood of $(x,y)$ is unchanged if spectral patterns in the neighborhood have a variety. That is, $k$ has different values even if the heat map is homogeneous. In this sense, $k$ is more informative than entropy and perplexity.

Different anatomical regions showed different entropy and perplexity (Fig.~\ref{fig3}). In normal kidneys, cortex contains more histological structures including proximal tubules, glomeruli, cortical distal tubules and interstitial structures, while medulla contains interstitial structures, medullary tubules, and corticomedullary tubules \cite{Martin-Saiz21}. Our detected $m/z$ range ($550$-$950$) were mainly lipids. It is known that the lipid expression of histological structures within the cortex is similar unlike the lipid expression within the medulla \cite{Martin-Saiz21}, which may explain the inconsistency of entropy and perplexity in Fig.~1.

The fact that the entropy and perplexity in the medulla area were both lower than those in the cortex area in Fig.~\ref{fig3} implies less complexity of biomolecular and chemical information in the medulla area. The major lipids in the cortex are phospholipids, whereas the medulla is dominated by neutral lipids \cite{Druilhet-etal78}. Phospholipids are composed of a head group, a glycerol backbone and fatty acid chains \cite{Drescher-vanHoogevest20}. Neutral lipids, on the other hand, contain only a glycerol backbone and fatty acid chains \cite{Athenstaedt-Daum06}. This might explain low-entropy spots in the medulla area.

When mass spectra drastically vary in space (e.g., when different high intensity peaks appear in neighboring spots), entropy grows by coarse-graining. In this case perplexity rapidly increases and $k$ at this spot becomes large. By contrast, if mass spectra are spatially more or less similar, entropy changes mildly by coarse-graining and $k$ is small.

In Fig.~\ref{fig5}, $k$ is large (red and yellow spots) around the vein and artery area in the kidney. The H\&E staining shows blood vessels. Indeed, pixels inside and outside the blood vessels have quite different mass spectra. On the other hand, $k$ is small (blue spots) in the medulla area. This implies that the region of medulla is relatively homogeneous.

\section{Conclusions}
\label{concl}

We have proposed the use of perplexity and introduced $k$ in coarse-graining. We found that the heat map of $k$ reveals new structures which are not clearly visible in the Shannon-entropy heat map. Although the use of peaks in the mass spectrum as a distribution is not yet establised, the spatial distribution of $k$ will help characterize biological tissues.

Since experimental data show the power-law behavior of the exponential of perplexity in (\ref{PPeps}), it was possible to produce heat maps of $k$ for kidneys. These heat maps elucidate structures of the Shannon entropy while the information was not clearly extracted with fractal dimensions.

\begin{acknowledgments}
This research was funded by MEXT Project for promoting public utilization of advanced research infrastructure (Imaging Platform), grant number JPMXS0410300220, AMED, grant number JP20gm0910004, JSPS KAKENHI, grant number JP18H05268, and the HUSM Grant-in-Aid, grant number 1013511. L.X. owes her sincere thanks to the Uehara Memorial Foundation Research Fellowship Program to support her study in Japan. M.M. acknowledges support from JST PRESTO Grant Number JPMJPR2027.
\end{acknowledgments}




\end{document}